\documentclass[a4paper,12pt]{article}
\usepackage[utf8]{inputenc}
\usepackage{cancel}
\usepackage{ulem}
\usepackage{amsfonts}
\usepackage{amssymb}
\usepackage{graphicx}
\usepackage{amsmath}
\usepackage{enumerate}
\usepackage{mathtools}
\usepackage{subfig}
\usepackage{color}
\usepackage{tikz}\usetikzlibrary{calc}
\usepackage{float}
\usepackage{here}
\usepackage{cite}
\usepackage{mathrsfs}
\usepackage{float,epsfig}
\usepackage{dcolumn}
\usepackage{tabularx}
\usepackage{multirow}
\usepackage{graphicx}
\usepackage{bm}
\usepackage{amsmath,amssymb,amsthm}
\usepackage[colorlinks=true,linkcolor=blue,citecolor=red]{hyperref}
\newcommand{\be}{\begin{equation}}
\newcommand{\ee}{\end{equation}}
\newcommand{\bea}{\setlength\arraycolsep{2pt} \begin{eqnarray}}
\newcommand{\eea}{\end{eqnarray}}

\def\0{{\sst{(0)}}}
\def\1{{\sst{(1)}}}
\def\2{{\sst{(2)}}}
\def\3{{\sst{(3)}}}
\def\4{{\sst{(4)}}}
\def\5{{\sst{(5)}}}
\def\6{{\sst{(6)}}}
\def\7{{\sst{(7)}}}
\def\8{{\sst{(8)}}}
\def\sst#1{{\scriptscriptstyle #1}}

\makeatletter \@addtoreset{equation}{section}

\usepackage{multirow}
\usepackage{tikz}

\begin{document}
%

\title{\normalsize
{\bf \Large	On  5D Black Brane Stabilities from  M-theory on   Three Parameter Calabi-Yau  Threefolds}}
\author{ \small A. Belhaj$^{1}$\footnote{a-belhaj@um5r.ac.ma}, H. Belmahi$^{1}$\thanks{hajar-belmahi@um5.ac.ma}, A. Bouhouch$^{1}$\thanks{abderrahim.bouhouch@um5r.ac.ma},  S. E. Ennadifi$^{2}$\thanks{Ennadifis@gmail.com},
\footnote{Authors in alphabetical order.}
	\hspace*{-8pt} \\
	{\small $^1$ D\'{e}partement de physique, \'Equipe des Sciences de la mati\`ere et du rayonnement,
		ESMaR}\\ {\small   Facult\'e des Sciences, Universit\'e Mohammed V de Rabat,  Rabat, Morocco} \\
	{\small $^{2}$  LHEP-MS, Facult\'e des Sciences, Universit\'e Mohammed V de Rabat,  Rabat, Morocco }
 }

 \maketitle

	\begin{abstract}
{\noindent}  In this work, we reconsider the study of 5D black branes in
M-theory compactifications by means of  $\mathcal{N}=2$ supergravity formalism.
Precisely, we provide a model relaying on a three parameter Calabi-Yau
manifold in the $\mathbb{P}^{1}\times\mathbb{P}^{1}\times\mathbb{P}^{2}$
projective space factorization, referred to as economical model. First, we
investigate the stability of 5D BPS and non-BPS black holes obtained from
wrapped M2-branes on non-holomorphic two-cycles in such a Calabi-Yau
manifold.  Then, we approach the stability of 5D black strings derived from
wrapped M5-branes on non-holomorphic four-cycles. Among others, we find
various stable and unstable black brane solutions depending on the charge
regions of the involved moduli space.

\textbf{Keywords}: 5D $N=2$ supergravity formalism, Black holes, Black
strings, Calabi-Yau manifolds, Stability Behaviors. 
\end{abstract}
 \newpage
\tableofcontents

%

\newpage

\newpage

\section{Introduction}

Black holes are considered as the most captivating and enigmatic entities in
the cosmos. Their studies have been essential in advancing the understanding
of the fundamental laws of the universe \cite{1,2,3,4,5,6}. The physical
properties of such objects have been dealt with by investigating both the
thermodynamic and the optical behaviors. By exploiting the Anti de Sitter
spaces, the thermodynamics of various black holes has been approached by
computing and analyzing the relevant quantities needed to discuss certain
aspects including the stability and the phase transitions \cite%
{7,8,9,9MOU,10,11,12}. Alternatively, the optical aspect has been extensively
investigated by examining concepts treating the light behaviors around black
holes \cite{13,130,131,132, 133,134,14,15,16,17,18,19,20,20H,20CH,20S}. Corroborated
studies of the black holes in non-trivial theories have been conducted by
means of observational findings obtained by Event Horizon Telescope (EHT)
international collaborations \cite{21,22,23,24}. These investigations have
been extended to black holes in high-energy theories. Concretely, the black
hole behaviors in type IIB superstring and M-theory have been exploited. In
superstring and M-theory scenarios, black holes in arbitrary dimensions have
been investigated using different methods and approaches based on analytical
and numerical techniques \cite{25,26,27,28,280}. In M-theory, for instance, the
first investigation of black holes has been successfully conducted by
computing the Bekenstein-Hawking entropy via the M2-branes moving inside the
Calabi-Yau (CY) threefold (3-fold) geometries with SU(3) holonomy group \cite{29,30,31,32}. These CY geometries have been constructed using various
methods explored in different investigation directions depending on desired
scenarios. It has been observed that the most studied ones are orbifolds,
toric hypersurface Calabi-Yau threefolds (THCY), and complete intersection
Calabi-Yau threefolds  (CICY)  in projective spaces. In fact, these methods complement
each other to give a unified picture of the compactification mechanism
extending the Kaluza-Kein scenario.  

Recently, the thermodynamic and the optical aspects of the black objects in
connection with Anti-de-Sitter (AdS) geometries have been explored by using
the brane physics. It has been remarked that the brane number has been
considered as a relevant parameter to examine the associated physical
behaviors \cite{33,34,35,36,36MOU}. Precisely, the size and the shape of the black
hole shadows have been studied by varying such a number. For certain brane
number values, it has been found that the black hole shadows in M-theory
exhibit non-trivial forms with different sizes including the  cardioid  curve 
geometries \cite{34}.

More recently, higher-dimensional black holes in supersymmetric M-theory
compactifications have been reconsidered via the attractor mechanism results 
\cite{360,361,362}. Particularly, the 5D black holes and the black strings
have been approached using the compactfication mechanism on CY geometries 
\cite{37,38,39,40,41,42}. In these activities, the BPS and the non-BPS black
object states have been constructed by exploiting the 5D $\mathcal{N}=2$
supergravity formalism developed in string theory activities and related
issues. More precisely, black holes, for instance, are raised by wrapping M2
branes on non-holomorphic 2-cycles of the CY three-folds while the black
string configurations are obtained by wrapping M5 branes on non-holomorphic
dual 4-cycles inside such CY geometries. These cycles are controlled by a
real number $h^{1,1}$ describing the dimension of the K\"{a}hler moduli
space of these geometries needed to determine the associated 5D spectrum
originated from M-theory. A close examination shows that two-dimensional K%
\"{a}hler moduli spaces of CY geometries, associated with $h^{1,1}=2$, have
been studied by discussing the stability behaviors of such black brane
objects. These studies have been elaborated by computing a scalar quantity
called the recombination factor $R$. A priori, stable and unstable brane
solutions could appear corresponding to $R<1$ and $R>1$, respectively \cite%
{37}. Later on, a three-dimensional K\"{a}hler moduli space has been also
approached by considering  toric geometry realizations of CY manifolds. The 5D
BPS and the non-BPS black brane solutions exhibiting stable and stable
behaviors have been derived from THCY  using numerical studies \cite{39}.

The aim of this work is to join these activities by proposing a
three-parameter CICY  model for black
brane stability investigations using the $\mathcal{N}=2$ 5D supergravity
formalism. Precisely, we present a M-theory compactification model relaying
on the $\mathbb{P}^{1}\times \mathbb{P}^{1}\times \mathbb{P}^{2}$ projective
space factorization with $h^{1,1}=3$, referred to as \textit{economical model%
}. Computing the effective scalar potential, we first find the 5D BPS and
the non-BPS black hole solutions and then examine their stability behaviors
from wrapped M2-branes on non-holomorphic two-cycles by providing a general
expression for the recombination factor. Using local variable
factorizations, we find stable and unstable solutions in different regions
of the black hole moduli space. After that, we approach 5D black string
solutions and examine their stability behaviors from wrapped M5-branes on
non-holomorphic four-cycles in the economical model. Among others, we find
various stable and unstable black brane solutions depending on the charge
regions of the moduli space.

The organization of this work is as follows. In section 2, we present a
concise discussion on 5D black objects from M-theory on CY 3-folds. In
section 3, we investigate the behaviors of 5D BPS and non-BPS black holes
from M2-branes wrapping 2-cycles on a three parameter   CICY  3-fold via an
economical model involving three ordinary projective spaces. In section 4,
we approach the 5D BPS and non-BPS black strings from such a CY 3-fold. In
the last section, we provide concluding remarks and open questions.

\section{5D black branes in M-theory CY compactifications}

In this section, we present a concise discussion on the 5D black holes and
the black strings using the $\mathcal{N}=2$ supergravity formalism explored
in the string theory compactification mechanism. In particular, these
objects can be obtained from M-theory on CY three-folds using M-brane
physics. The CY geometries have been explored first in superstring theory
compactifications to generate models with minimal supercharges in 4D
space-times \cite{43,44,45,46,47,48}. Alternatively, they have been used
also in the compactification of M-theory to provide 5D space-time
supersymmetric models. At lower-energy limits, this theory is described by
11D supergravity with $\mathcal{N}=2$ supersymmetry producing certain
superstring models in 10D via the dimension reduction scenarios \cite{29,30,480,481}%
. Roughly, this 11D theory contains also two solitonic objects called M2 and
M5-branes. Using the compactification mechanism, these objects have been
exploited to build 5D black holes and black strings in CY three-fold
compactifications, respectively. To approach such black brane solutions, one
uses the 5D $\mathcal{N}=2$ supergravity formalism via the following
Maxwell-Einstein action 
\begin{equation}
S=\frac{1}{2\kappa _{5}^{2}}\int d^{5}x\left( R\star \mathbb{I}%
-G_{IJ}dt^{I}\wedge \star dt^{J}-G_{IJ}F^{I}\wedge \star F^{J}-\frac{1}{6}%
C_{IJK}F^{I}\wedge F^{J}\wedge A^{K}\right) 
\end{equation}%
involving $h^{1,1}$ vector multiples denoted by the index $I$ containing  the  scalar   K\"{a}hler  moduli   $t_{I}$. This Hodge
number counts the K\"{a}hler deformations of 2-cycles of the CY three-folds
being involved in the black brane building models from M-theory
compactifications. In this action, the symmetric tensor $G_{IJ}$ represents
the moduli space metric being a needed relevant quantity to express the
effective scalar potential. Roughly speaking, 5D black holes carry $q^{I}$
electric charges under the $U(1)^{h^{1,1}}$ gauge symmetry. These charges
can be exploited to determine the associated effective potential using the
charge space metric. In this way, the black hole potential can be expressed
as follows 
\begin{equation}
V_{eff}^{e}=G^{IJ}q_{I}q_{J}.  \label{Vbh}
\end{equation}%
Considering the dual version corresponding to the black strings with the
magnetic charges $p^{I}$, the associated scalar potential takes the
following form 
\begin{equation}
V_{eff}^{m}=4G_{IJ}p^{I}p^{J}.  \label{vm}
\end{equation}%
It has been shown that the moduli space metric $G_{IJ}$ can be expressed in
terms of the CY three-fold volume via the relation 
\begin{equation}
G_{IJ}=-\frac{1}{2}\partial _{I}\partial _{J}\log (\mathcal{V}),
\end{equation}%
where the volume $\mathcal{V}$ reads as 
\begin{equation}
\mathcal{V}=\frac{1}{3!}C_{IJK}t^{I}t^{J}t^{K}.
\end{equation}%
In this equation, the tensor $C_{IJK}$ represents the intersecting numbers.
These numbers could be fixed once the CY geometries are constructed. It has
been remarked that many CY manifold constructions have been elaborated. The
most popular ones are toric varieties and intersecting hypersurfaces in
non-trivial projective space fibrations. For instance,  certain CY geometries can be
considered as CICYs in products of projective spaces called the ambient
space $\mathcal{A}=\mathbb{P}^{n_{1}}\times ...\times \mathbb{P}^{n_{m}}$.
It is recalled that a $n_{\ell }$-dimensional ordinary projective space $%
\mathbb{P}^{n_{\ell }}$ is defined by considering the following
identification 
\begin{equation}
(z_{1},\ldots ,z_{n_{\ell }+1})\sim (\lambda z_{1},\ldots ,\lambda
z_{n_{\ell }+1})
\end{equation}%
where $(z_{1},\ldots ,z_{n_{\ell }+1})$ are the homogeneous complex
coordinates and $\lambda $ is a nonzero complex scalar. Concretely, CICYs
can be specified by a configuration matrix encoding the information of the
embedding space and the homogeneous degrees of the involved polynomials. In
general, it is a $m\times k$ matrix which can be represented by the
following scheme

\begin{equation}
\label{matrix}
	\mbox{CY}_3(\mathcal{A})= \left[\begin{array}{cccc}
	\mathbb{P}^{n_{1}} \\ 
	\vdots \\
	\mathbb{P}^{n_{m}}
	\end{array} \right|\left| \begin{array}{ccc}
	d_{1}^{1}& \ldots& d_{k}^{1}\\ 
	\vdots& \ddots & \vdots \\ 
	d_{1}^{m}& \ldots & d_{k}^{m}
	\end{array} \right]_{\chi}^{h^{1,1}, h^{2,1}}
	\end{equation}
subject to the following constraints 
\begin{equation}
\sum\limits_{r=1}^{m}n_{r}-k=3,\qquad n_{r}+1=\sum\limits_{\alpha
=1}^{k}d_{\alpha }^{r}.
\end{equation}%
The numerical quantities $h^{1,1}$ and $h^{2,1}$ are the fundamental Hodge
numbers controlling the K\"{a}hler and the complex structure deformations of
CY three-folds, respectively. The number $\chi =2(h^{1,1}-h^{1,2})$
denotes the Euler number which is a topological invariant. These Hodge
numbers provide the associated physical spectrum content involving vector
multiplets and hypermultiplets. It is noted in passing that the Hodge number 
$h^{2,1}$ describing the shape deformations is not relevant in the black
brane activities. These deformations are linked to hypermultiplets being
evinced in the present investigation. In connection with 5D black branes
from M-theory, many models of two parameter CY three-folds have been
investigated. More details can be found in \cite{37,38,40,41,42}. These
studies have been extended to a model based on a three parameter CY
considered as THCYs where stable and non-stable solutions have been
approached \cite{39}. Here, we reconsider such investigations by providing a
three parameter CY model via CICY scenarios in products of certain ordinary
projective spaces, hereinafter referred to as {\it economical model}.

\section{Black holes from M-theory on a CICY with $h^{1,1}=3$}

In this section, we reconsider the study of 5D black holes from three
parameter CY threef-folds by proposing an economical model of CICY
scenarios. In particular, we investigate 5D black holes from a CICY with $%
h^{1,1}=3$. These theories involve $U(1)\times U(1)\times U(1)$ gauge fields
producing solutions possessing three electric charges which can be denoted
by the charge triplet $(q_1,q_2,q_3).$ In this M-theory scenario, these
charges correspond to M2-branes wrapped on non-holomorphic 2-cycles of the
proposed CY three-fold associated with the K\"{a}hler moduli space. The
manifold constructed here is considered as the intersection of the
hypersurfaces in a product of the three ordinary projective spaces called
the ambient space given by 
\begin{equation}
\mathcal{A}=\mathbb{P}^{1}\times\mathbb{P}^{1}\times\mathbb{P}^{2}
\end{equation}
where one has used $n_1=n_2=1$ and $n_3=2.$ In this way, we can provide an
economical model to study the 5D black holes via a configuration matrix, which
comprises the data on the intersection of the hypersurfaces, with
three-degrees $(2,2,3)$ as the common zero locus of three homogeneous
polynomials in the embedded space $\mathcal{A}$. In this way, the above
configuration matrix reduces to 

\begin{equation}
\text{CY}_3(\mathcal{A})= \left[\begin{array}{c}
\mathbb{P}^{1} \\ 
\mathbb{P}^{1} \\ 
\mathbb{P}^{2}
\end{array}
\right|\left|  
\begin{array}{c}
2 \\ 
2 \\ 
3
\end{array}
\right]_{-144}^{3,75}
\end{equation} 

The crucial geometric data of such a CY  three-fold  are the intersection numbers needed
to determine the relevant black hole quantities including the effective
potential. For this three parameter CY  three-fold, these numbers are found to be 
\begin{eqnarray}
C_{133}&=&C_{233}=2,\,\,\, C_{123}=3  \notag \\
C_{111}&=&C_{222}=C_{333}=C_{112}=C_{122}=C_{113}=C_{223}=0
\end{eqnarray}
which provide the volume of the proposed CY being written as follows 
\begin{equation}
\mathcal{V}=t_3(t_1 (3 t_2+t_3)+t_2 t_3).
\end{equation}
The associated central charge is expressed as 
\begin{equation}
Z_{e}=q_1 t_1+ q_2 t_2+ q_3 t_3.
\end{equation}
Using the relation (\ref{Vbh}), we obtain the effective potential which
reads as 
\begin{equation}
V^e_{eff}=\frac{2F(t_1,t_2,t_3)}{{9 t_1 (3 t_2 + t_3)+ 3 t_3 (3 t_2 + 4 t_3)}%
},
\end{equation}
where one has used 
\begin{align*}
F(t_1,t_2,t_3)&=6 q_2 q_3 t_3^2 \left( t_2 t_3 - t_1 (3 t_2 + t_3)+ 3 q_3^2
t_3^2 (3 t_1 (3 t_2 + t_3)+ t_3 (3 t_2 + 2 t_3)\right) \\
&+ q_1^2 \left( 2 t_2 t_3^3 + 9 t_1^3(3 t_2 + t_3)+ 2 t_1 t_3^2 (6 t_2 +
t_3) + 3 t_1^2 t_3 (9 t_2 + 4 t_3)\right) \\
&- 2 q_1 t_3^2 \left( 2 q_2 (t_1 + t_2) t_3 + 3 q_3 (3 t_1 t_2 - t_1 t_3 +
t_2 t_3)\right) \\
&+ q_2^2 \left( t_2 t_3 (9 t_2^2 + 12 t_2 t_3 + 2 t_3^2) + t_1 (27 t_2^3 +
27 t_2^2 t_3 + 12 t_2 t_3^2 + 2 t_3^3)\right) .
\end{align*}
At this level, we would like to provide certain comments on such  a black hole
effective potential. First, the latter is a weight-2 function of the
homogeneous variables $t_i$. The second comment concerns the $\mathbb{Z}_2$
symmetry invariance of such a potential sending $q_i$ to $-q_i$. This
symmetry will provide symmetric cone charges of the possible black hole
solutions. This could be exploited to validate the obtained black hole
regions.

 To determine  and examine  5D BPS and non-BPS black hole solutions, however,  the  local coordinates of the black hole  moduli space will be exploited. In  this investigation, the  following local variables
\begin{equation}
\rho=\frac{q_1}{q_3}, \qquad  \sigma=\frac{q_2}{q_3}, \qquad x=\frac{t_1}{t_3},\qquad y=\frac{t_{2}}{t_3},
\end{equation}
 will be considered   where $ q_I$ and $t_I$ denote the homogeneous coordinates of the  5D black hole  moduli space associated with the charges and the  K\"{a}hler  deformations of the  three parameter  CY three-fold,    respectively.
\subsection{BPS and non-BPS  black hole   solutions}
We start by  studying the BPS solutions of such an economical  model. Indeed, we determine  the critical point  equations of the  three parameter CY geometry   by solving the constraints 
\begin{equation}
q_I-2 \tau_I Z_e=0,\,\,\,I=1,2,3,
\end{equation}
where  one  has used $
\tau_I= \frac{1}{2}  C_{IJK}    t^J t^K $.
In this way, we obtain a  system of three equations involving the  local variables 
\begin{equation}
\begin{array}{rl}
\left(\rho -2 t_3^3 (3 y+1) (\rho  x+\sigma  y+1)\right) &= 0 \\
 \left(\sigma -2 t_3^3 (3 x+1) (\rho  x+\sigma  y+1)\right)&=0\\
  \left(1-2t_3^3 (x (3 y+2)+2 y) (\rho  x+\sigma  y+1)\right)&=0.
\end{array}
\end{equation}
To find the BPS black hole charges, we should solve the above equations subject  to the  CY volume constraint.  We  can express   the local charge variables in terms of the geometric ones associated with the CY K\"{a}hler  moduli by eliminating   the   homogeneous variable $ t_3$.  After calculations, we get
\begin{equation}
\rho =\frac{3 y+1}{2 x+(3 x) y+2 y},\,\,\,\sigma =\frac{3 x+1}{2 x+(3 x) y+2 y},
\label{bbs}
\end{equation}
  exhibiting  a nice mapping symmetry
\begin{eqnarray}
\sigma & \leftrightarrow &  \rho   \nonumber \\
x  &\leftrightarrow&  y.
\end{eqnarray}
The above  equations can be handled to get $x$ and $y$ as functions of $\rho$ and $\sigma$.  Approaching  such  equations,   we find  four solutions
\begin{eqnarray}
x_\pm &=& \frac{\pm\sqrt{\rho ^2 +\sigma ^2+14 \rho  \sigma -6 \rho -6 \sigma +9}  -\sigma -3 \rho +3}{6 \rho}\\
y_\pm &=& \frac{\pm\sqrt{\rho ^2 +\sigma ^2+14 \rho  \sigma -6 \rho -6 \sigma +9}-\rho -3 \sigma +3}{6 \sigma}.
\end{eqnarray}
To determine the possible electric  charge regions, we should impose constraints on the geometric  local variables  of  the  black hole moduli space. A priori, there are many ways to do that.  However,  we follow here  a method based on the separation of $x$ and $y$  local geometric variables in order to illustrate  such  behaviors in two dimensional plane. In Fig.(\ref{F1}), we  show the allowed charge  regions for BPS  black hole solutions. 

\begin{figure}[h!]
\label{F1}
\begin{center}
	\includegraphics[scale=0.7]{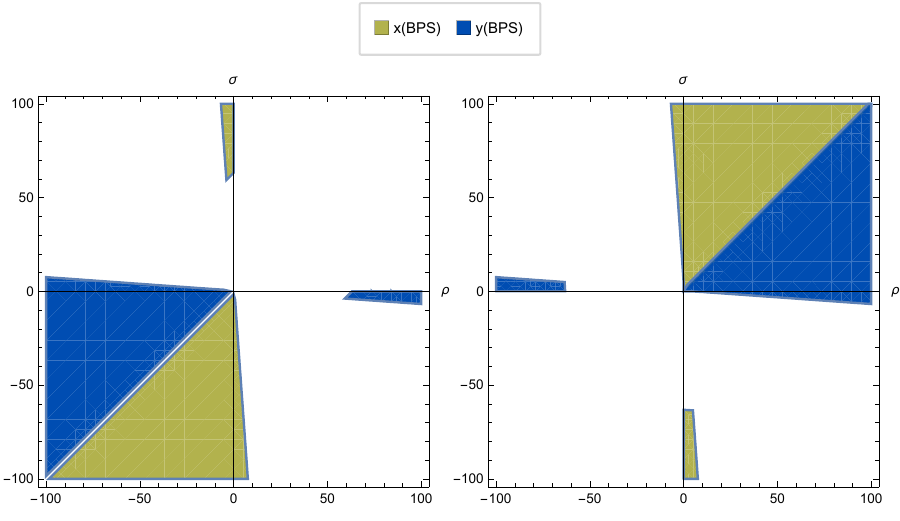}
	\caption{{\it \footnotesize  Left panel:  Electric charge regions for  non BPS black hole sates of the pair $(x_+, y_+) $.   Right panel:  Electric charge regions for  non BPS black hole sates of the pair $(x_-, y_-) $.  }}
\label{F1}
\end{center}
\end{figure}
In the left  and the right panels, we illustrate the regions  of  large black holes for the pair $(x_+, y_+) $  and  $(x_-, y_-) $ solutions, respectively. These  configurations    provide only half cone black hole solutions.  The  white regions correspond to  the absence of  large black hole solutions.   The  small regions  of both panels can be viewed as an overlapping behaviors of  $x_{\pm}$ and $y_{\pm}$   BPS solutions.  They can be   combined with large ones    to recover the symmetric cone behaviors of the BPS  black hole solutions.

Using the results of \cite{37}, we  can approach 
the entropy of the BPS black holes  using   the  following   formula 
\begin{equation}
S_{BPS}=\frac{ 2 \pi}{3\sqrt{3}} |Z_e|^{3/2}.
\end{equation}
In terms of the  black hole moduli space, this entropy  can be expressed as   
\begin{equation}
S_{BPS}=\frac{ 2 \pi}{3\sqrt{3}}| t_3 q_3|^{3/2} |x\rho+y\sigma+1|^{3/2}
\end{equation}	
where $t_3$  reads  as 
\begin{equation}
t_3=\frac{1}{\sqrt[3]{6} \sqrt[3]{3 x y+x+y}}
\end{equation}
being  fixed via the  volume of  the proposed CY 3-fold. In this way, the entropy takes the following general form 
\begin{equation}
S_{BPS}=\frac{\sqrt{2} \pi }{9} \left| \frac{q_3 (x \rho +y \sigma +1)}{\sqrt[3]{3 y x+x+y}}\right| ^{3/2}
\end{equation}
involving  the  cubic root charge behaviors. This quantity  could be exploited    to approach   other  thermodynamic quantities.  Indeed,  it has been  shown that  there are many  extended    entropy forms in the   black hole investigations  \cite{OS1,OS2}.  Motivated by such generalized entropies, we  could discuss   the  thermodynamic behaviors  of  these 5D black hole solutions  by  focusing on  the relevant quantity being the temperature.   Indeed,  we could  identify the obtained entropy $ S_{BPS}$  with the   Brow one  denoted by   $S_{B}$ 
\begin{equation}
S_{B}=\left(\frac{A}{A_{p_{1}}}\right)^{1+\frac{\Delta}{2}},
\end{equation}
where   $ A_{p_{1}}=4G $ is the  the Planck area and one   has considered  $ A=4\pi r_{h}^{2} $.     $\Delta$    is  a   dimensionless  parameter  which  denotes   the quantum gravity deformation.  Taking a   maximal quantum deformation  conditioned  by   $\Delta=1$,   the corresponding  event horizon radius  is found to be 
\begin{equation}
r_{h}= \frac{1}{2}\left(\frac{1}{9} \sqrt{2} \pi \right)^{1/3}\left| \frac{q_3 (x \rho +y \sigma +1)}{\sqrt[3]{3 y x+x+y}}\right|^{1/2}.
\end{equation}
After calculations, we  obtain the   Hawking  temperature
\begin{equation}
T_{H}= \left(\frac{3\sqrt{3}}{(2\pi)^{3}|q_3(x\rho+y\sigma+1)|}\right)^{1/2}.
\end{equation}
For generic regions of the moduli space,  the obtained temperature exhibits inverse  square root  charge behaviors.
 
	 

In order to determine the   non-BPS solutions, we use the Lagrange multipliers  by means of  the constraint $g_{\cal V}=t_3(t_1 (3 t_2+t_3)+t_2 t_3)$. 
Using this equation
\begin{equation}
\label{defVeff}
\dfrac{D_{I}V^e_{eff}}{D_{J}V^e_{eff}}=\dfrac{D_{I}g_{\cal V}}{D_{J}g_{\cal V}},\,\,\,\,\,I,J=1,2,3,
\end{equation}
we obtain  two algebraic equations  given by 
\begin{align*}
&27 x^4 (3 y+2) (\rho +3 \rho  y)^2+36 \rho ^2 x^3 (3 y+1) (3 y+2)^2-3 x^2 (-30 \rho ^2-4 \rho  \sigma +6 \rho +2 \sigma ^2\\
&-6 \sigma +81 \sigma ^2 y^4-27 y^3 \left(5 \rho ^2+6 \rho -5 \sigma ^2+6 \sigma -9\right)-27 y^2 (13 \rho ^2+2 \rho  \sigma -3 \sigma ^2+6 \sigma -9)\\
&-3 y \left(61 \rho ^2+2 \rho  (5 \sigma -6)-7 \sigma ^2+18 \sigma -27\right)+9)+x (4 \rho ^2
-8 \rho  \sigma +12 \rho +4 \sigma ^2-12 \sigma +81 \sigma ^2 y^4\\
&+54 y^3 \left(\rho ^2+4 \rho  \sigma +6 \rho +3 \sigma ^2-6 \sigma -9\right)+9 y^2 (33 \rho ^2+18 \rho  \sigma +6 \rho+13 \sigma ^2-36 \sigma -63)\\
&+12 y \left(11 \rho ^2+2 \rho  \sigma -3 \rho +3 \sigma ^2-9 \sigma -18\right)-27)+54 \sigma ^2 y^4+9 y^3 \left(8 \rho  \sigma +6 \rho +8 \sigma ^2-18 \sigma -9\right)\\
& +6 y^2 \left(5 \rho ^2+6 \rho  \sigma +3 \rho +5 \sigma ^2-27 \sigma -18\right)+y \left(4 \rho ^2-8 \rho  \sigma +36 \rho +4 \sigma ^2-36 \sigma -99\right)-24=0
\end{align*}
together  with 
\begin{align*}
&(x (9 y+3)+3 y+4) (\rho +3 \rho  x-\sigma  (3 y+1)) (3 x^2 (\rho +3 \rho  y)+\rho  x (6 y+3)+\sigma  x (3 y+1)^2\\
&+3 \sigma  y^2+y (\rho +3 \sigma )+2)=0,
\end{align*} 
where  one  has used $D_I= \partial_I-\frac{2}{3{\cal V}} \tau_I$.
A close examination shows that it  is  possible to solve these  two algebraic equations  in order to obtain   the non-BPS  black hole charges in terms of the geometric local variables $x$ and $y$. Precisely, we find four black hole  solutions
\begin{equation}
\label{1}
\rho = \frac{3 y+1}{3 x y+2 x+2 y},\,\,\,\sigma = \frac{3 x+1}{3 x y+2 x+2 y}
\end{equation}
\begin{equation}
\label{2}
\rho = \frac{x (9 y+3)+3 y-1}{9 x^2 y+x (9 y+2)+2 y},\,\,\,\sigma= -\frac{(3 x+1)^2}{9 x^2 y+x (9 y+2)+2 y}
\end{equation}
\begin{equation}
\label{3}
\rho = -\frac{(3 y+1)^2}{x \left(9 y^2+9 y+2\right)+2 y},\,\,\,\sigma = \frac{x (9 y+3)+3 y-1}{x \left(9 y^2+9 y+2\right)+2 y}
\end{equation}
\begin{equation}
\label{4}
\rho = (3 y+1)S(x,y),\,\,\,\sigma = (3 x+1) S(x,y)
\end{equation}
where the function $S(x,y)$ is expressed as follows
\begin{equation}
\label{5}
S(x,y)=-\frac{ \left(9 x \left(6 y^2+5 y+1\right)+9 (3 x y+x)^2+9 y^2+9 y+8\right)}{f(x,y)}
\end{equation}
with
\begin{align}
f(x,y)&=27 x^3 y (3 y+1)^2+3 x^2 \left(54 y^3+135 y^2+69 y+10\right)+x \left(27 y^3+207 y^2+132 y+16\right)\notag\\
&+2 y (15 y+8).
\end{align}
The first solution of  (\ref{1}) is associated with  the BPS  charge solutions  obtained in (\ref{bbs}) being examined in the previous  part. The remaining ones   do not satisfy the  BPS  equations providing  non-BPS  object solutions.   We   first  examine the second and the third one   in detail.   Indeed,  an inspection  shows that we have two solution types,  which will  be discussed in detail.  The first type of  solutions are  given by 
\begin{equation}
x_{1,2}= \frac{\pm\sqrt{\tau}-3 \rho -\sigma +3}{6 \rho },\qquad y_{1,2}=\frac{\mp\sqrt{\tau}+5 \rho -\sigma -15}{18 \sigma }
\end{equation} 
where one has used $\tau=\rho ^2+14 \rho  \sigma -6 \rho +\sigma ^2-6 \sigma +9$.
However, the second ones  are  given by 
\begin{equation}
x_{3,4}=\frac{\pm\sqrt{\tau}-\rho -5 \sigma +15}{18 \rho }, \qquad y_{3,4}= \frac{\mp\sqrt{\tau}+\rho -3 \sigma +3}{6 \sigma }.
\end{equation}
These solutions are constrained by  the K\"{a}hler  cone conditions $x_\pm^{1,2},y_\pm^{1,2}>0$.

\begin{figure}[h!]
\begin{center}
\begin{tikzpicture}[scale=0.2,text centered]
\hspace{ 0cm}

\node[] at (30,28) {\small  \includegraphics[scale=1.1]{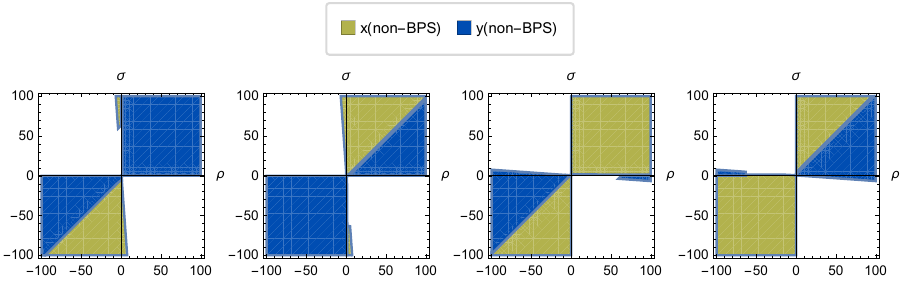}};
\node[] at (0,14) {\small  (a)};
\node[] at (20,14) {\small   (b)};
\node[] at (41,14) {\small (c)};
\node[] at (62,14) {\small   (d)};

\end{tikzpicture}
\end{center}
\caption{{\it \footnotesize   Allowed electric charge regions for   the non-BPS black hole states of the ($x,y$) pairs.  (a):$(x_{1},y_{1})$, (b):$(x_{2},y_{2})$,  (c): $(x_{3},y_{3})$, (d) :$(x_{4},y_{4})$}}
\label{F2}
\end{figure}
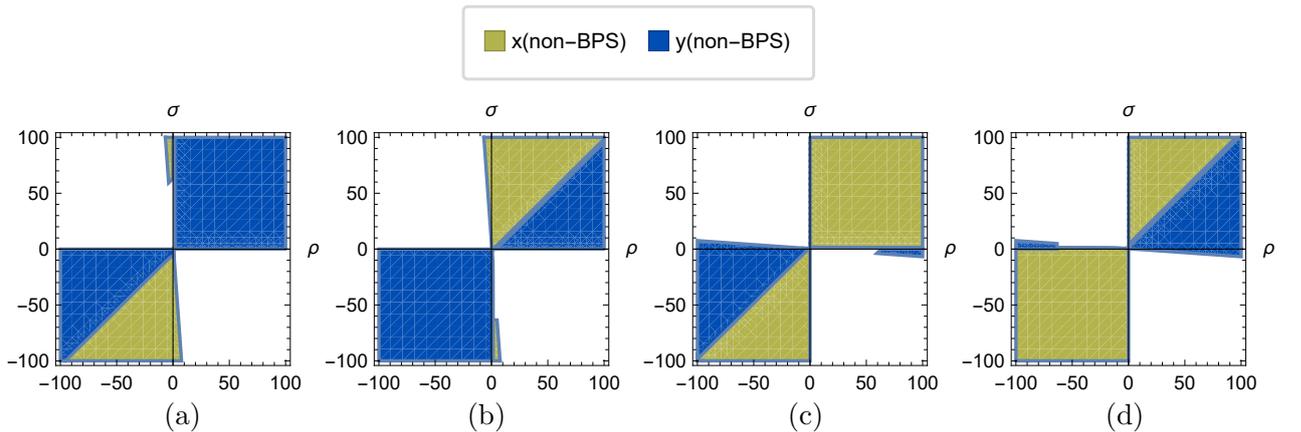

In Fig.(\ref{F2}), we illustrate  the allowed charge regions in  the half cone configurations   for the four solutions of the non-BPS black holes.

Similar to the previous solutions, we observe overlapping configurations  via  small and large  regions, which could be combined  to recover the symmetric behavior of the  K\"{a}hler cones.   It has been remarked that  the   Fig.(\ref{F2})(a) and (b) can be grouped to generate  the possible symmetric  cone for the first solution  family.   However,     Fig.(\ref{F2})(c) and (d) can be grouped to provide the second family of  the possible symmetric cones.  These graphical representations could  be viewed as   extended  configurations  of two parameter CY manifold  where only  one symmetric cone appears  for each non-PBS solution.

 Now, we come back  to  the  last solution  given by  (\ref{3}). To approach  this solution,  we should solve   the constraint 
\begin{equation}
f(x,y)=0
\end{equation}
being an algebraic  equation  of degree 3 in   $(x,y)$  variables.  It has been observed that it is not an  easy task   to  express such local  variables in terms of  the $(\rho, \sigma)$ charge ratios.  Numerical computations show that 
\begin{equation}
x=\dfrac{f(y)}{54 A(y) y (3 y+1)^2}
\end{equation}
where one has used 
\begin{align}
f(y)&=-A \left(108 y^3+270 y^2+138 y-2^{2/3}A+20\right)+1458 \sqrt[3]{2} y^6-7290 \sqrt[3]{2} y^5\\ 
&+7128 \sqrt[3]{2} y^4+18846 \sqrt[3]{2} y^3+10818 \sqrt[3]{2} y^2+2472 \sqrt[3]{2} y+200 \sqrt[3]{2}\nonumber
\end{align}
and
 \begin{align}
A^3(y)&=-2000-37080 y-275292 y^2-1027188 y^3-1911681 y^4-1232010 y^5+887922 y^6  \nonumber\\&+997272 y^7-295245 y^8+39366 y^9
+(27y^2 (3 y+1)^6 (150903 y^8-1194102 y^7+\nonumber\\&3606363 y^6+1161540 y^5-6135588 y^4
-5901120 y^3-2196864 y^2-380928 y-25600))^{1/2}.
\end{align}
The  above equation involves a negative  trivial pole  $y^t_p$   and non-trivial $y^{nt}$  ones verifying  $A(y^{nt}_p)=0$. Near  non-trivial poles,   certain zeros  could  be obtained by  solving the 
\begin{equation}
1458  y^6-7290  y^5
+7128  y^4+18846  y^3+10818  y^2+2472 y+200=0
\end{equation}
A examination shows that these zeros  involve negative  values  being removed in the present discussion due the  K\"{a}hler cone conditions.

\subsection{Stability analysis}
 Now, we move to  examine the stability behaviors of the  non-BPS black holes,  since the BPS  ones  remain stable due the supersymmetry  aspect.   To do so, we need  to  determine  the recombination factor of each solution. According to \cite{37}, the recombination factor of non-BPS black holes     can be obtained  via the relation 
 \begin{equation}
 R=\frac{\sqrt{{V^{cr}_{eff}}}}{{t_I} | {q_I}| }
 \end{equation}
where  $V^{cr}_{eff}$ represents   the effective  potential values of the black holes  at the critical points. A close examination shows that we could provide a general expression. Indeed, 
the computation gives
\begin{equation}
R=\frac{1}{2(1+x|\rho| +y|\sigma|)}\sqrt{\frac{3\zeta}{3 x^2+3 x (6 y+1)+9 y^2+3 y+1}}
\end{equation} 
where one has used
\begin{align}
\zeta&=x^4 \left(15 \rho ^2-6 \rho  \sigma +\sigma ^2\right)+2 x^3 (6 y+1) \left(9 \rho ^2-4 \rho  \sigma +\sigma ^2\right)\\ &+2 x^2 \left(-2 \sigma +3 \rho ^2 \left(27 y^2+9 y+1\right)-2 \rho  \sigma  \left(27 y^2+9 y+1\right)+\sigma ^2 \left(27 y^2+9 y+1\right)+6\right)\notag \\ &+x \left(\rho ^2-2 \rho  (\sigma -2)\right)+\sigma ^2-4 \sigma +36 y^3 \left(3 \rho ^2-4 \rho  \sigma +3 \sigma ^2\right)+18 y^2 \left(3 \rho ^2-4 \rho  \sigma +3 \sigma ^2\right) \notag
 \\&+4 y \left(3 \rho ^2-4 \rho  \sigma +3 \left(\sigma ^2-2 \sigma +6\right)\right)+12+9 y^4 \left(3 \rho ^2-6 \rho  \sigma +5 \sigma ^2\right)+6 y^3 \left(3 \rho ^2-6 \rho  \sigma +5 \sigma ^2\right)\notag \\ &+2 y^2 \left(3 \rho ^2-6 \rho  \sigma +5 \sigma ^2-6 \sigma +18\right)+y \left(\rho ^2-2 \rho  \sigma +\sigma ^2+12\right)+2. \notag
\end{align}
For the four solutions presented previously, we analyze  the stability behaviors on the allowed charge regions by considering separately the regions of ($x=0, y>0$) and ($x=0, y>0$). In what follows,  we discuss  numerically the   corresponding recombination factors.   Starting with  the first solution $(x_1, y_1)$, six  branches have been obtained.  In particular, we find two class of  three  regions. The first class is associated with  ($x_1=0, y_1>0$)  while the second one corresponds to  three  other ones  with    ($x_1>0, y_1=0$).  These separated  three regions are   illustrated in Table (1).  

\begin{table}[!ht]
  \begin{center}
  \begin{tabular}{ |c|c|c|c| } 
\hline
 & \scriptsize  Region 1 &\scriptsize  Region 2 & \scriptsize  Region 3 \\
 
\hline
\multirow{2}{4em}{\tiny  $x_{1}=0,y_1>0$} &\tiny $\rho<\frac{\sigma +6}{2}-\frac{1}{2} \sqrt{\sigma ^2+6 \sigma }$ &\tiny  $\frac{1}{2} \sqrt{\sigma ^2+6 \sigma }+\frac{\sigma +6}{2}<\rho \leq -2 \sqrt{3} \sqrt{4 \sigma ^2-3 \sigma }-7 \sigma +3$ &\tiny  $\rho >\frac{\sigma +6}{2}-\frac{1}{2} \sqrt{\sigma ^2+6 \sigma }$ \\ 
& \tiny  $\sigma \leq -\frac{75}{8}$ &\tiny  $-\frac{75}{8}<\sigma \leq-6 $&\tiny  $\sigma >0$  \\ 
\hline

\multirow{2}{4em}{\tiny $x_{1}>0,y_1=0$}  & \tiny $2 \sqrt{3} \sqrt{4 \sigma ^2-3 \sigma }-7 \sigma \leq \rho <0 $  & \tiny $\frac{1}{2} (2 \sigma +3)<\rho \leq -2 \sqrt{3} \sqrt{4 \sigma ^2-3 \sigma }-7 \sigma +3 $ & \tiny $ 0<\rho \leq -2 \sqrt{3} \sqrt{4 \sigma ^2-3 \sigma }-7 \sigma +3$ \\
& \tiny $ \sigma >3$  & \tiny $-\frac{3}{2}<\sigma <-\frac{3}{8} $ & \tiny $ \sigma \leq -\frac{3}{2}$ \\
\hline
\end{tabular}
 \caption{\it \footnotesize Classification  of  electric charge regions of  the non-BPS black holes associated with $(x_1, y_1)$ pair solutions.  }
\label{tab1}
\end{center}
\end{table}

\begin{figure}[h!]
\begin{center}
\begin{tikzpicture}[scale=0.2,text centered]
\hspace{ 0cm}
\node[] at (1,28) {\small  \includegraphics[width=5.5cm, height=4.5cm]{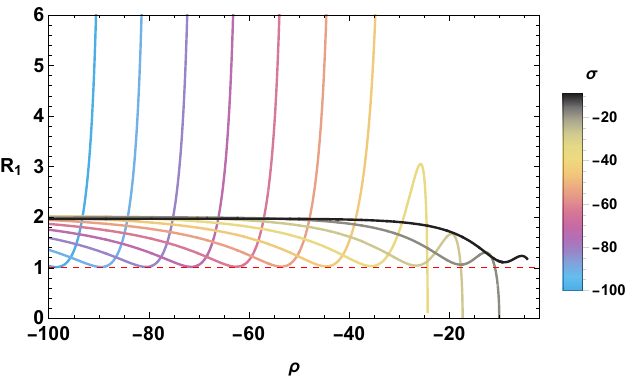}};
\node[] at (30,28) {\small  \includegraphics[width=5.5cm, height=4.5cm]{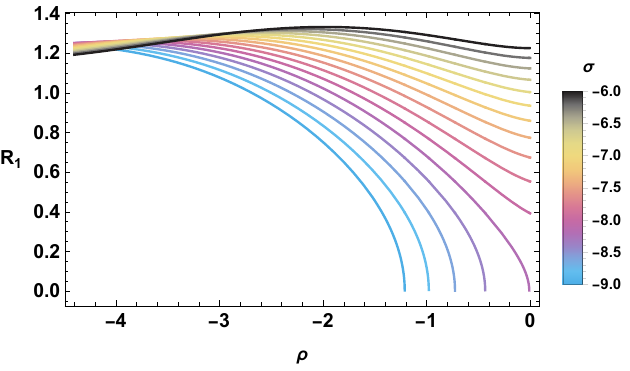}};
\node[] at (60,28) {\small  \includegraphics[width=5.5cm, height=4.5cm]{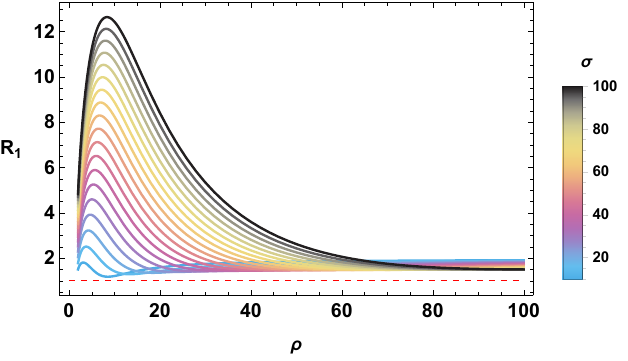}};

\node[] at (1,42) {\scriptsize    Region 1};
\node[] at (30,42) {\scriptsize    Region 2};
\node[] at (60,42) {\scriptsize    Region 3};

\end{tikzpicture}
\caption{{{\it \footnotesize Recombination factor  for $(x_{1}=0,y_1>0)$  in the three allowed charge regions. }}}
\label{S1x}
\end{center}
\end{figure}

To compute and  discuss   the corresponding expressions of the  recombination factor,   we deal with  the three  branches of  the non-BPS black holes of  each  case   of   the $(x_1, y_1)$  pair solution.
 In Fig.(\ref{S1x}),  we  illustrate the behaviors  of the recombination factor  $R_1$ associated with $(x_1=0, y_1>0)$   by considering the three allowed  charge regions.

Since it has been shown  that a solution could be stable if this factor is more than one,  it follows from this  figure that  we have found  stable and unstable black hole configurations.
In the region 3,  for instance,   we observe that  the recombination factor is always greater than one.  This reveals that  the 
associated  non-BPS black holes  are all unstable.    They  decay to BPS and anti BPS states  matching with the results of \cite{37}. For  the regions 1 and 2,  however,  we  find    stable and unstable black hole configurations.   Considering the region 2 and taking a fixed value of  $\sigma$,   the value of the recombination factor  decreases by increasing  $\rho$.  Moreover, we observe that  the black holes become stable after a  certain critical value.

\begin{figure}[h!]
\begin{center}
\begin{tikzpicture}[scale=0.2,text centered]
\hspace{ 0cm}
\node[] at (1,28) {\small  \includegraphics[width=5.5cm, height=4.5cm]{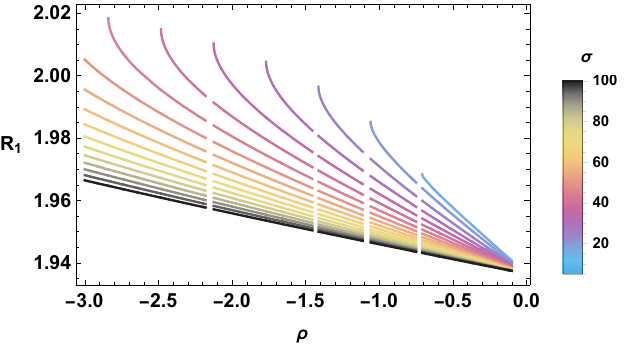}};
\node[] at (30,28) {\small  \includegraphics[width=5.5cm, height=4.5cm]{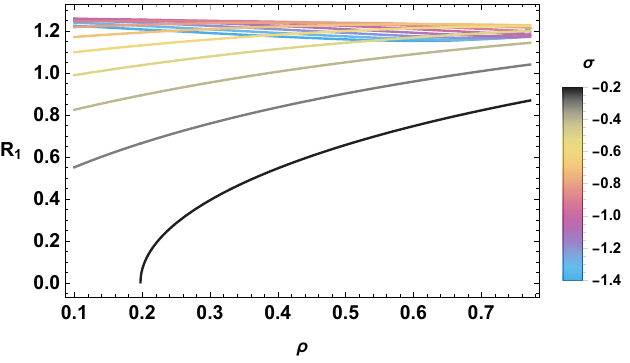}};
\node[] at (60,28) {\small  \includegraphics[width=5.5cm, height=4.5cm]{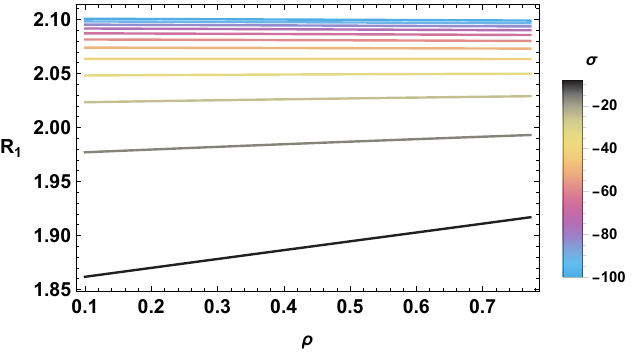}};

\node[] at (1,42) {\scriptsize    Region 1};
\node[] at (30,42) {\scriptsize    Region 2};
\node[] at (60,42) {\scriptsize    Region 3};

\end{tikzpicture}
\caption{{{ \it \footnotesize Recombination factor  for $(x_{1}>0,y_1=0)$  in the three allowed charge regions. }}}
\label{S1y}
\end{center}
\end{figure}

Taking the second case  where ($x_{1}>0,y_1=0$) of the first solution, the behaviors of the  corresponding recombination factor $R_1$  are illustrated  in  Fig.(\ref{S1y}).

In the regions 1 and 3,   the recombination factor is always greater than one showing that the corresponding  non-BPS black holes  are all unstable.    They  decay to BPS and anti BPS states  matching with the results of \cite{37}.  In the region 2,   the  value of   recombination factor increases  by increasing $\rho$ for the allowed values of the charge variable $\sigma$.  For certain critical points,  the black holes become unstable.

Moving now to approach the second solution of  the $(x_2, y_2)$ pair.  The allowed charge regions are presented in  Table 2. 
 \begin{table}[!ht]
  \begin{center}
  \begin{tabular}{ |c|c|c|c| } 
\hline
 & \scriptsize  Region 1 &\scriptsize  Region 2 & \scriptsize  Region 3 \\
 
\hline
\multirow{2}{4em}{\tiny  $x_{2}=0,y_2>0$} &\tiny $\rho <\frac{1}{2} \sqrt{\sigma ^2+6 \sigma }+\frac{\sigma +6}{2}$ &\tiny  $\rho \leq -2 \sqrt{3} \sqrt{4 \sigma ^2-3 \sigma }-7 \sigma +3 $ & \tiny  $ \rho >\frac{1}{2} \sqrt{\sigma ^2+6 \sigma }+\frac{\sigma +6}{2}$ \\ 
& \tiny  $\sigma \leq -\frac{75}{8}$ &\tiny  $ -\frac{75}{8}<\sigma <0 $&\tiny  $\sigma >0$  \\ 
\hline

\multirow{2}{4em}{\tiny $x_{2}>0,y_2=0$}  & \tiny $ 0<\rho \leq -2 \sqrt{3} \sqrt{4 \sigma ^2-3 \sigma }-7 \sigma +3$  & \tiny $0<\rho <\frac{1}{2} (2 \sigma +3) $ & \tiny $2 \sqrt{3} \sqrt{4 \sigma ^2-3 \sigma }-7 \sigma +3\leq \rho <0 $ \\
& \tiny $ \sigma <-\frac{3}{8}$  & \tiny $-\frac{3}{8}\leq \sigma \leq 3 $ & \tiny $ \sigma >3$ \\
\hline
\end{tabular}
 \caption{\it \footnotesize Classification  of  electric charge regions of the  non-BPS black hole solutions  corresponding to  $(x_2, y_2)$ pair solutions.}
\label{tab2}
\end{center}
\end{table}

\begin{figure}[h!]
\begin{center}
\begin{tikzpicture}[scale=0.2,text centered]
\hspace{ 0cm}
\node[] at (1,28) {\small  \includegraphics[width=5.5cm, height=4.5cm]{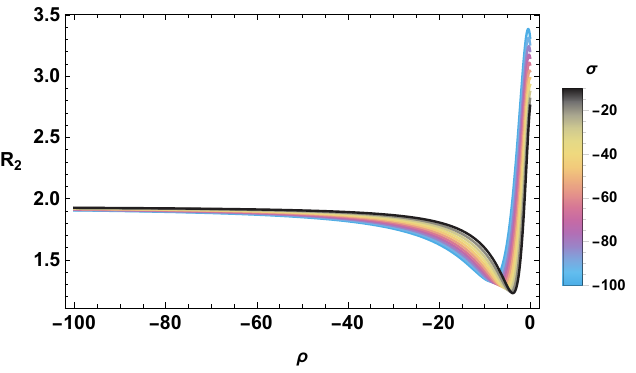}};
\node[] at (30,28) {\small  \includegraphics[width=5.5cm, height=4.5cm]{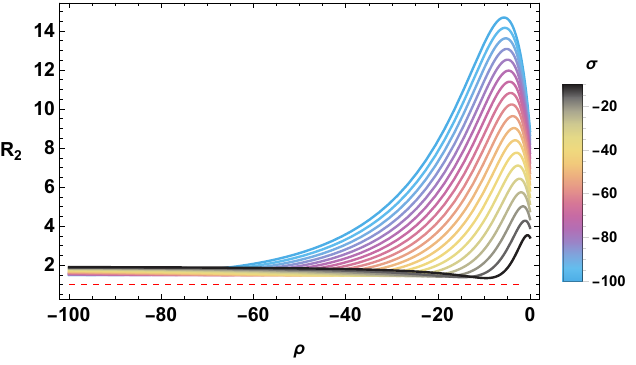}};
\node[] at (60,28) {\small  \includegraphics[width=5.5cm, height=4.5cm]{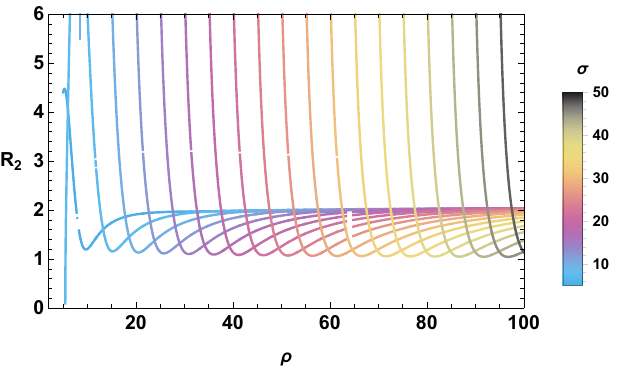}};

\node[] at (1,42) {\scriptsize    Region 1};
\node[] at (30,42) {\scriptsize    Region 2};
\node[] at (60,42) {\scriptsize    Region 3};

\end{tikzpicture}
\caption{{{ \it \footnotesize Recombination factor  for $x_{2}=0,y_2>0$  in the three allowed charge regions. }}}
\label{s211}
\end{center}
\end{figure}

In Fig.(\ref{s211}), we present the behaviors of the recombination factor  $R_2$ for the ($x_{2}=0,y_2>0$) pair solution.

\begin{figure}[h!]
\begin{center}
\begin{tikzpicture}[scale=0.2,text centered]
\hspace{ 0cm}
\node[] at (1,28) {\small  \includegraphics[width=5.5cm, height=4.5cm]{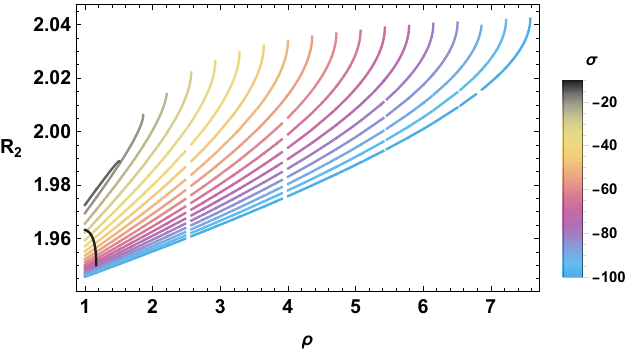}};
\node[] at (30,28) {\small  \includegraphics[width=5.5cm, height=4.5cm]{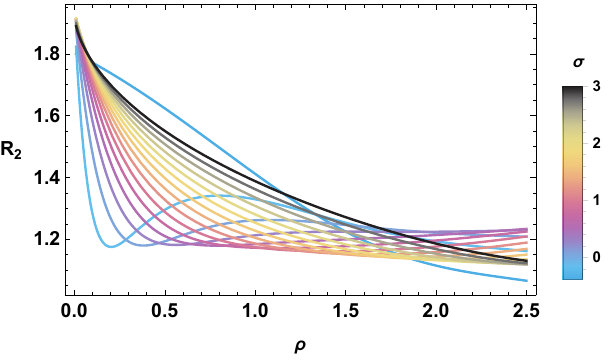}};
\node[] at (60,28) {\small  \includegraphics[width=5.5cm, height=4.5cm]{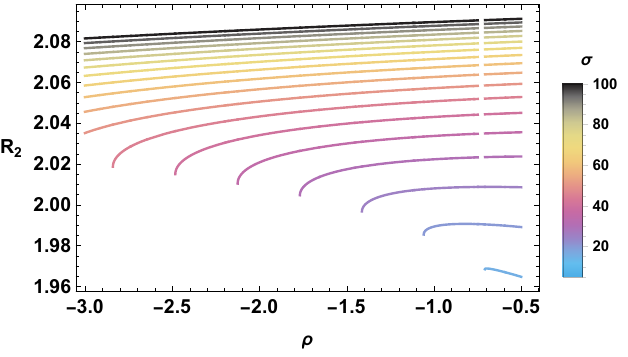}};

\node[] at (1,42) {\scriptsize    Region 1};
\node[] at (30,42) {\scriptsize    Region 2};
\node[] at (60,42) {\scriptsize    Region 3};

\end{tikzpicture}
\caption{{{ \it \footnotesize Recombination factor  for $x_{2}>0,y_2=0$  in the three allowed charge regions. }}}
\label{S2X}
\end{center}
\end{figure}

For all three regions,  it has been observed that   the recombination factor is always greater than  one 
 showing  that the  black
hole solutions  remain unstable.  
The  recombination factor $R_2$  corresponding to  the  second case where ($x_{2}>0,y_2=0$) is depicted in Fig.(\ref{S2X}) for the three regions. 

For all three regions,  similarly,   we find that  the  black
hole solutions  are  unstable which  prefer   to decay into the associated 
BPS and anti-BPS  brane states.

 The third solution  provides  the  allowed regions presented in Table 3. 
  \begin{table}[!ht]
\begin{center}
  \begin{tabular}{ |c|c|c|c| } 
\hline
 & \scriptsize  Region 1 &\scriptsize  Region 2 & \scriptsize  Region 3 \\
 
\hline
\multirow{2}{4em}{\tiny  $x_{3}=0,y_3>0$} &\tiny $\rho <\frac{1}{2} (2 \sigma -3) $ &\tiny  $ \rho \geq 2 \sqrt{3} \sqrt{4 \sigma ^2-3 \sigma }-7 \sigma +3$ & \tiny  $\rho \leq -2 \sqrt{3} \sqrt{4 \sigma ^2-3 \sigma }-7 \sigma +3 $ \\ 
& \tiny  $\sigma <0 $ &\tiny  $ \sigma <0 $&\tiny  $\sigma >\frac{9}{8} $  \\ 
\hline

\multirow{2}{4em}{\tiny $x_{3}>0,y_3=0$}  & \tiny $\frac{2 \sigma ^2-12 \sigma +18}{2 \sigma -3}<\rho <0 $  & \tiny $ \rho >0 $ & \tiny $  $ \\
& \tiny $\sigma \leq \frac{3}{4} $  & \tiny $\sigma >3 $ & \tiny $  $ \\
\hline
\end{tabular}
 \caption{\it \footnotesize Classification  of  electric charge regions of non-BPS  solutions associated with $(x_3, y_3)$ pair solutions.}
\label{tab3}
\end{center}
\end{table}

\begin{figure}[h!]
\begin{center}
\begin{tikzpicture}[scale=0.2,text centered]
\hspace{ 0cm}
\node[] at (1,28) {\small  \includegraphics[width=5.5cm, height=4.5cm]{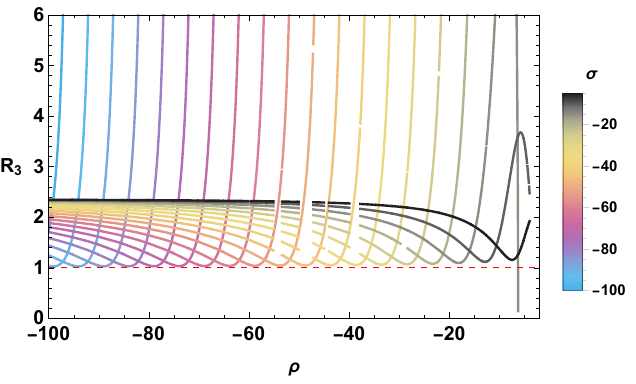}};
\node[] at (30,28) {\small  \includegraphics[width=5.5cm, height=4.5cm]{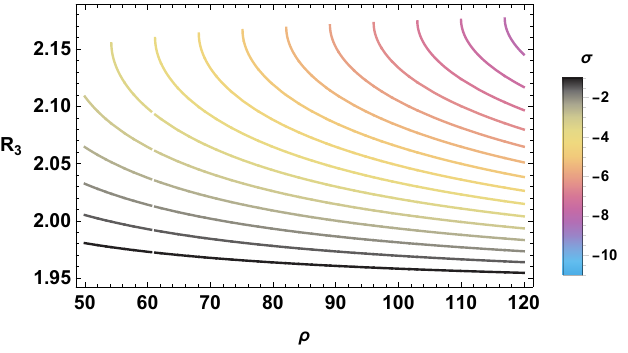}};
\node[] at (60,28) {\small  \includegraphics[width=5.5cm, height=4.5cm]{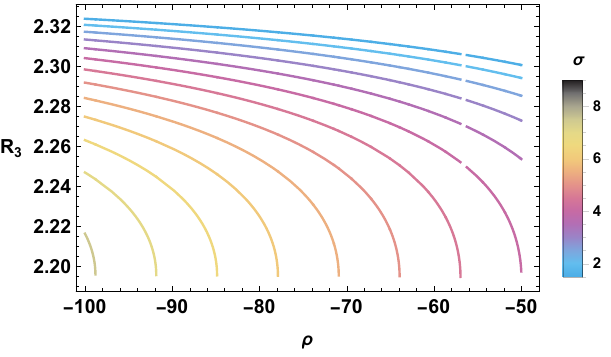}};

\node[] at (1,42) {\scriptsize    Region 1};
\node[] at (30,42) {\scriptsize    Region 2};
\node[] at (60,42) {\scriptsize    Region 3};

\end{tikzpicture}
\caption{{{\it \footnotesize Recombination factor  for $x_{3}=0,y_3>0$  in the three allowed charge regions. }}}
\label{s3Y}
\end{center}
\end{figure}
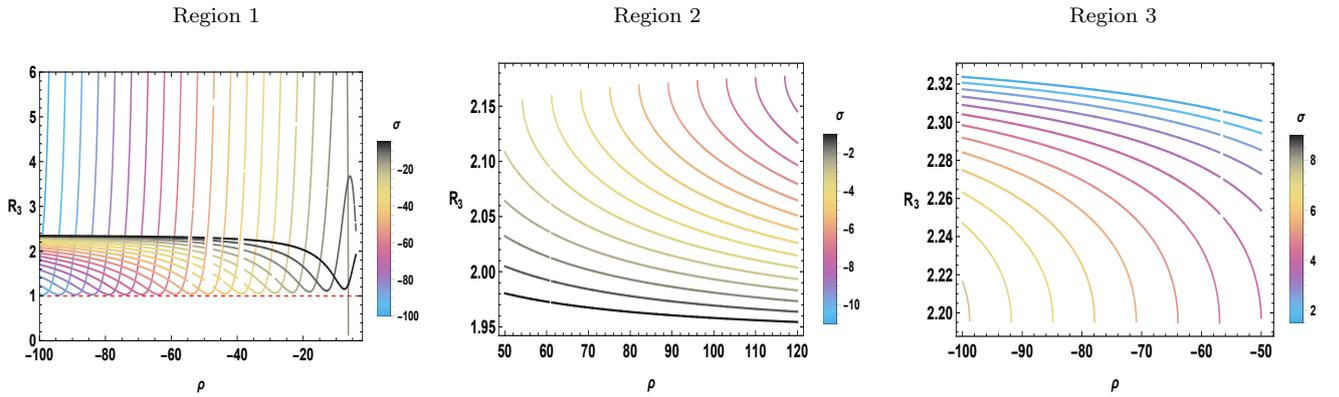

Considering the case ($x_{3}=0,y_3>0$), the behaviors of the recombination factor $R_3$ of the corresponding three regions  are  illustrated in Fig.(\ref{s3Y}).

\begin{figure}[h!]
\begin{center}
\begin{tikzpicture}[scale=0.2,text centered]
\hspace{ 0cm}
\node[] at (1,28) {\small  \includegraphics[width=5.5cm, height=4.5cm]{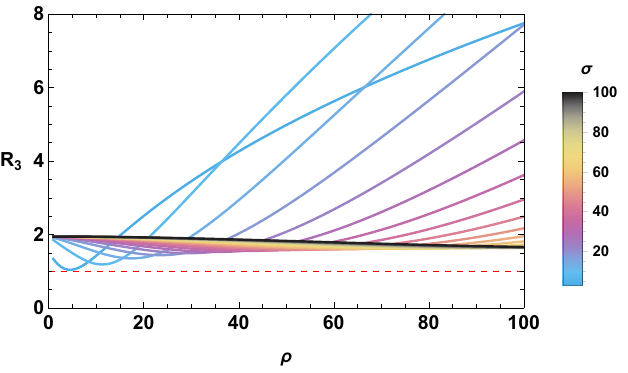}};
\node[] at (30,28) {\small  \includegraphics[width=5.5cm, height=4.5cm]{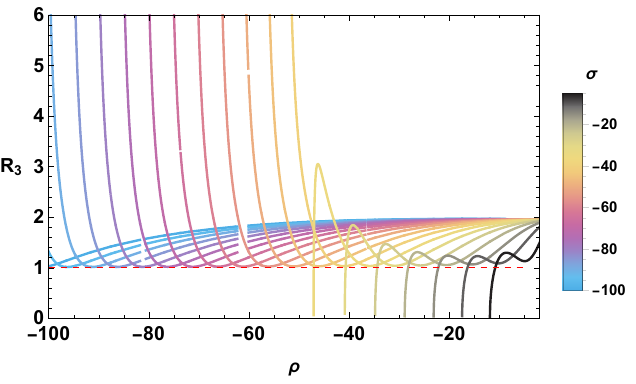}};

\node[] at (1,42) {\scriptsize    Region 1};
\node[] at (30,42) {\scriptsize    Region 2};

\end{tikzpicture}
\caption{{{\it \footnotesize Recombination factor  for $x_{3}>0$ and  $y_3=0$  in the three allowed charge regions. }}}
\label{s3x}
\end{center}
\end{figure}

In all regions,  it  has been remarked  that the  non-BPS black holes remain unstable.  For the second case corresponding to ($x_{3}>0,y_3=0$),  we find only two regions.  For such regions,  we  approach   numerically 
 the stability  behaviors  being presented in Fig.(\ref{s3x}). 
 In the region 1,   the  value of    the recombination factor increases  by increasing $\rho$ for the allowed values of the charge variable $\sigma$.  The associated values are  always greater than one  revealing that the corresponding  non-BPS black holes  are all unstable  decaying  to BPS and anti BPS brane  states.   For certain critical points in the region 2,  the black holes become stable.

\begin{figure}[h!]
\begin{center}
\begin{tikzpicture}[scale=0.2,text centered]
\hspace{ 0cm}
\node[] at (1,28) {\small  \includegraphics[width=5.5cm, height=4.5cm]{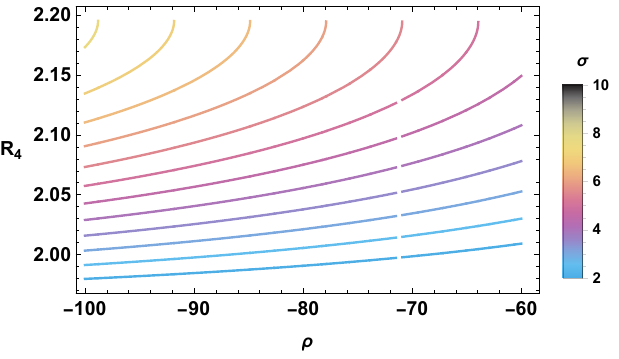}};
\node[] at (30,28) {\small  \includegraphics[width=5.5cm, height=4.5cm]{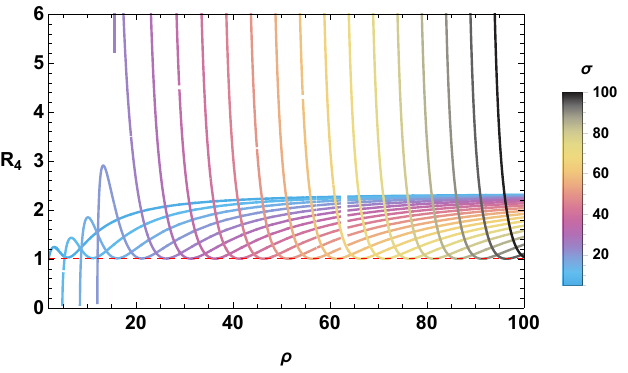}};
\node[] at (60,28) {\small  \includegraphics[width=5.5cm, height=4.5cm]{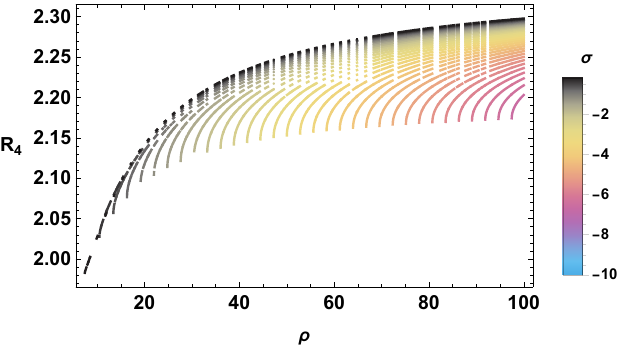}};

\node[] at (1,42) {\scriptsize    Region 1};
\node[] at (30,42) {\scriptsize    Region 2};
\node[] at (60,42) {\scriptsize    Region 3};

\end{tikzpicture}
\caption{{{\it \footnotesize  Recombination factor  of  non-BPS black holes   for $x_{4}=0$  and $y_4>0$  in the three allowed charge regions. }}}
\label{s4Y}
\end{center}
\end{figure}
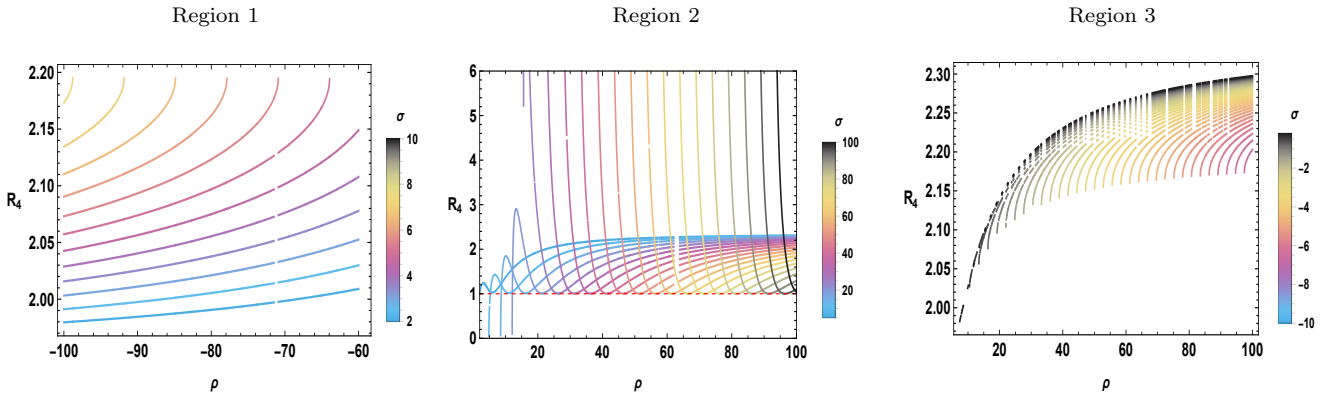

Moving now to inspect  the stability of the last  pair solution. The  corresponding  possible  charge   regions are   given in Table 4. 
  \begin{table}[!ht]
\begin{center}
  \begin{tabular}{ |c|c|c|c| } 
\hline
 & \scriptsize  Region 1 &\scriptsize  Region 2 & \scriptsize  Region 3 \\
 
\hline
\multirow{2}{4em}{\tiny  $x_{4}=0,y_4>0$} &\tiny $\rho \leq -2 \sqrt{3} \sqrt{4 \sigma ^2-3 \sigma }-7 \sigma +3$ &\tiny  $ \rho >\frac{1}{2} (2 \sigma -3)$ & \tiny  $\rho \geq 2 \sqrt{3} \sqrt{4 \sigma ^2-3 \sigma }-7 \sigma +3 $ \\ 
& \tiny  $\sigma \geq \frac{9}{8} $ &\tiny  $\sigma \geq \frac{9}{8}  $&\tiny  $ \sigma <0$  \\ 
\hline

\multirow{2}{4em}{\tiny $x_{4}>0,y_4=0$}  & \tiny $\rho <0 $  & \tiny $ \rho <\frac{2 \sigma ^2-12 \sigma +18}{2 \sigma -3}$ & \tiny $  $ \\
& \tiny $ \sigma \leq \frac{3}{4}$  & \tiny $ \sigma >3$ & \tiny $  $ \\
\hline
\end{tabular}
 \caption{\it \footnotesize Classification  of  the electric charge regions of non-BPS  solutions associated with $(x_4, y_4)$ pair solutions.}
\label{tab3}
\end{center}
\end{table}
The  numerical computations of the first case ($x_{4}=0,y_4>0$)  are  presented in Fig.(\ref{s4Y}). 

\begin{figure}[h!]
\begin{center}
\begin{tikzpicture}[scale=0.2,text centered]
\hspace{ 0cm}
\node[] at (1,28) {\small  \includegraphics[width=5.5cm, height=4.5cm]{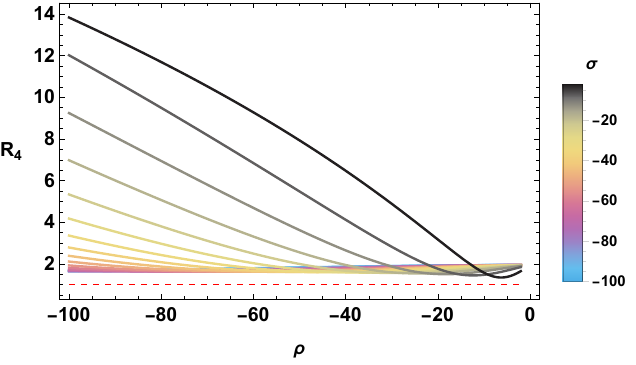}};
\node[] at (30,28) {\small  \includegraphics[width=5.5cm, height=4.5cm]{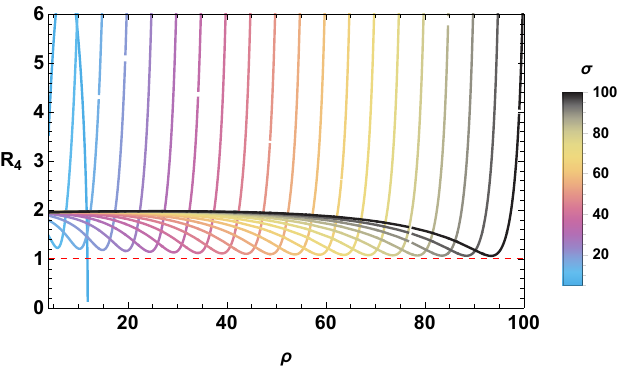}};

\node[] at (1,42) {\scriptsize    Region 1};
\node[] at (30,42) {\scriptsize    Region 2};

\end{tikzpicture}
\caption{{{\it \footnotesize Recombination factor of  non-BPS black holes    for $x_{4}>0$ and $y_4=0$  in the three allowed charge regions. }}}
\label{s4x}
\end{center}
\end{figure}

As the third solution,  in all regions, it  has been remarked  that the black holes remain unstable.   For the second case corresponding to  ($x_{4}>0,y_4=0$),  we obtain also  only two regions.  The associated stability  behaviors   are depicted   in Fig.(\ref{s4x}). 
In the region 1 and 2,   the  values of    the recombination factor   are  always greater than one   showing  that the associated   non-BPS black holes  are all unstable.   These states prefer to  decay to BPS and anti BPS  brane states.

\section{M-theory black strings from  a CICY with   $ h^{1,1}=3$}
In this section, we  investigate  the  5D black strings using  an  economical M-theory compactification in terms of three projective spaces. These  black object solutions can  be  derived  by    wrapping  M5-branes on  dual divisors  controlled by the K\"{a}hler moduli space of the  proposed CY manifold.  As  the previous model, we  study the  stability  examination  of   the black string states via  the corresponding   effective  potential  as a function of the M5-brane  magnetic  charges.   Indeed, we first  determine the  5D black  stringy solutions.  Then, we  examine their stabilities via the computation of the recombination factor. 
\subsection{Black string solutions}
We start by finding the black string solutions carrying $p^I$ magnetic charges where one has  $I=1,2,3$. To do so, we first  compute   the associated  effective potential   by considering  the proposed three parameter  CICY  manifold.   Using  similar techniques, this 
 effective potential  is found to be
\begin{eqnarray}
V^m_{eff}&=& \frac{1}{18}p_1^2 t_3^2 (3 t_2+t_3)^2+\frac{1}{9} p_1 p_2 t_3^4+\frac{1}{3} p_1 p_3 t_2^2 t_3^2+\frac{1}{18}p_2^2 t_3^2 (3 t_1+t_3)^2+\frac{1}{3} p_2 p_3 t_1^2 t_3^2 \notag\\
&+&\frac{1}{18} p_3^2 \left(t_1^2 \left(9 t_2^2+6 t_2 t_3+2 t_3^2\right)+2 t_1 t_2 t_3 (3 t_2+2 t_3)+2 t_2^2t_3^2\right).
\end{eqnarray}
To determine the   critical points,  the local variable of the moduli space  will be required. Indeed, they  take  the following ratio forms 
\begin{equation}
 \alpha=\frac{p_1}{p_3}, \qquad  \beta=\frac{p_2}{p_3}, \qquad  x=\frac{t_1}{t_3}, \qquad   y=\frac{t_{2}}{t_3}.
\end{equation}
Using (\ref{defVeff}) to provide an  extremization  mechanism with respect  to the  geometric local variables subject to the volume constraint, we find the following  two  equations of motion 
associated with  the scalar fields defining the black string moduli space which are 
 \begin{align}
&-2 (\alpha +\beta )^2+9 x^3 \left(-\beta  (3 \beta +2)+3 y^2+2 y\right)+x^2 \left(-36 \beta ^2-6 \beta +27 y^2+9 y+2\right)\notag\\
&+x \left(-3 \beta  (4 \alpha +5 \beta )+9 y^2+4 (3 \alpha +1) y\right)+\left(9 \alpha ^2+6 \alpha +2\right) y^2-3 \alpha ^2 y=0,
\end{align}
and
 \begin{align}
 (\alpha +\beta +3 \beta  x+x+3 \alpha  y+y) (-\alpha +\beta +3 \beta  x+x-(3 \alpha +1) y)=0.
 \end{align}
 Solving these scalar  equations of motion, we  can obtain the magnetic  black string configurations. They are organized as follows:\\
First  charge solution:
\begin{eqnarray}
\alpha = x,\qquad \beta = y
\end{eqnarray}
Second  charge solution: 
\begin{eqnarray}
\alpha = -\frac{3 x y+x+2 y}{3 y+1},\qquad \beta = y
\end{eqnarray}
Third   charge solution:
\begin{eqnarray}
\alpha = x,\qquad \beta = -\frac{3 x y+2 x+y}{3 x+1}.
\end{eqnarray}
Fourth  charge solution:
{\small 
\begin{eqnarray}
 \alpha  =  -\frac{3 (3 x y+x)^2+9 x y (3 y+1)+2 y (3 y+4)}{(3 y+1) (x (9 y+3)+3 y+8)},
\beta &=&-\frac{3 x^2 \left(9 y^2+9 y+2\right)+x \left(18 y^2+9 y+8\right)+3 y^2}{(3 x+1) (x (9 y+3)+3 y+8)}.\notag\\
\end{eqnarray}}
To  determine the allowed magnetic charge regions, we should express the local geometric variable  in terms of the local magnetic ratio   charges.  A careful scrutiny shows that  there are   two independent solutions  in terms of   $(x,y)$ pairs.    Combining these charge solutions equations, we  obtain two  pair solutions
\begin{eqnarray}
x&=& \alpha, \qquad  \qquad\qquad \qquad \; \;y= -\frac{3 \alpha  \beta +2 \alpha +\beta }{3 \alpha +1} \\
x&=&-\frac{3 \alpha  \beta +\alpha +2 \beta }{3 \beta +1}, \qquad y= \beta .
\end{eqnarray}
As in the black hole case,  these solutions exhibit a nice mapping symmetry
\begin{eqnarray}
\alpha & \leftrightarrow &  \beta   \nonumber \\
x  &\leftrightarrow&  y.
\end{eqnarray}

\begin{figure}[h!]
\begin{center}
	\includegraphics[scale=0.7]{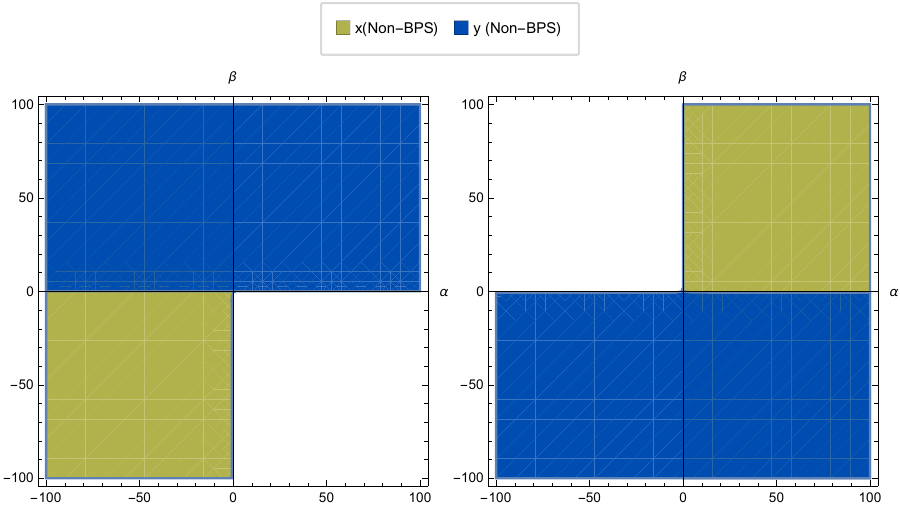}
	\caption{{\it \footnotesize   Magnetic charge regions for  non-BPS black string sates of the ($x,y$) pairs. }}
\label{F1bs}
\end{center}
\end{figure}

Examining the above  solutions, we could provide the possible magnetic charge regions. Using the previous techniques exploited for black holes, we illustrate such regions in Fig.(\ref{F1bs}) for non-BPS black strings.
As for black  holes, we find overlapping   regions  with symmetric  K\"{a}hler  cones.  This shows the existence of three  family  solutions  given by bleu and olive colors associated with the  two local geometric  variables.
\subsection{Stability scenarios}
To analyze the stability of such solutions, one needs to compute the recombination factor being   the ratio of the black string tension $T$  to that of the minimal  size of  the   corresponding piecewise calibrated divisor.  Using the local magnetic charge variable,  we   can get   such a factor   the 
  the  size  $ V_{D^{\cup}} $  which is   the minimum volume piecewise calibrated representative of the class $ [D] $ given by 
 \begin{equation} 
 D=p_{1} {\cal D}_1+p_{2}{\cal D}_2 +p_{3}{\cal D}_3
 \end{equation}
  where ${\cal D}_1$, ${\cal D}_2 $  and ${\cal D}_3$ are dual divisors  to    $ {\cal J}_1$,  ${\cal J}_2$ and  ${\cal J}_3$   being  the  K\"{a}hler  $(1,1)$-forms of    $\mathbb{P}^{1}$, $\mathbb{P}^{1}$ and  $\mathbb{P}^{2}$, respectively. In this way,  the recombination  factor  reads as 
 \begin{equation}
R=\frac{T}{ V_{D^{\cup}}}. 
\end{equation} 
In this equation,  one has $ V_{D^{\cup}}=\sum\limits _{I=1}^3A_{I}|p_{I}|$   where   $ A_{I}=C_{IJK}t^{J}t^{K}=2\tau_{I} $ describing the size of the divisors in  the proposed CY geometry.  After computations, we find 

\begin{equation}
V_{C^{\cup}}= |p_1| (6 t_2 t_3 + 2 t_3^2) +|p_2| (6 t_1 t_3 + 2 t_3^2) + 2 |p_3| ( 3 t_1 t_2 + 2 t_1 t_3 + 2 t_2 t_3).
\end{equation}
This leads to  the following general formula of the  black string  recombination factor  
{\small{
\begin{equation}
R=\frac{\sqrt{(\alpha +\beta )^2+x^2 \left(9 \beta ^2+6 \beta+2\right) +  y^2\left(9 \alpha ^2+6 \alpha +2\right)  + 6 (\beta ^2  x+  \alpha ^2 y)+6 (yx^2 + y^2x)+4 yx+  + 9 y^2 x^2 }}{6 \sqrt{2} (| \alpha | +| \beta | +3 x | \beta | +3 y | \alpha | +3 x y+2 x+2 y)}
\end{equation}}}
being invariant under the above  mapping symmetry. Considering the two solutions mentioned in the above subsection, the investigation  of the recombination factor for the allowed magnetic charge regions will be based on the branches illustrated in  Table 5.
  \begin{table}[!ht]
\begin{center}
  \begin{tabular}{ |c|c|c|c| } 
\hline
 & \scriptsize  Region 1 &\scriptsize  Region 2 & \scriptsize  Region 3 \\
 
\hline
\multirow{2}{5em}{\tiny  First solution} &\tiny $x=\alpha, y=-\frac{3 \alpha  \beta +2 \alpha +\beta }{3 \alpha +1} $ &\tiny  $ x=\alpha, y=0$ & \tiny  $ x=0,y=-\frac{3 \alpha  \beta +2 \alpha +\beta }{3 \alpha +1} $ \\ 
& \tiny  $\beta<0, \alpha>0 $ &\tiny  $ \beta>0, \alpha>0 $&\tiny  $\beta<\frac{-2}{3}, \alpha<-\frac{\beta }{3 \beta +2} $ \\ 
\hline

\multirow{2}{5em}{\tiny Second solution} &\tiny $x=-\frac{3 \alpha  \beta +\alpha +2 \beta }{3 \beta +1}, y=\beta $ &\tiny  $ x=0, y=\beta$ & \tiny  $ x=-\frac{3 \alpha  \beta +\alpha +2 \beta }{3 \beta +1},y=0 $ \\ 
& \tiny  $\beta>0, \alpha<0 $ &\tiny  $ \beta>0, \alpha>0 $&\tiny  $\beta<\frac{-1}{3}, \alpha<-\frac{2 \beta }{3 \beta +1} $ \\ 
\hline
\end{tabular}
 \caption{\it \footnotesize Classification  of  the magnetic charge regions.}
\label{tab3}
\end{center}
\end{table}

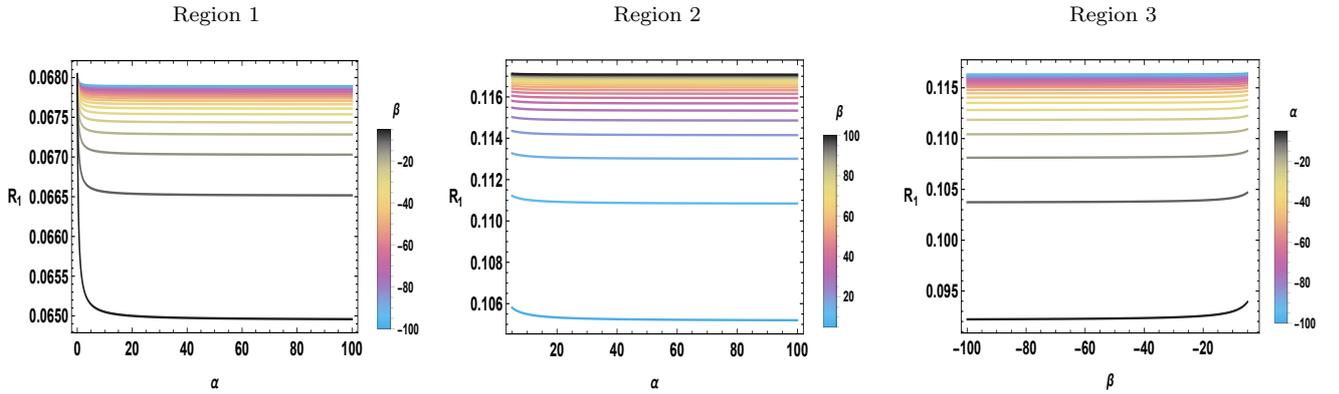
\begin{figure}[h!]
\begin{center}
\begin{tikzpicture}[scale=0.2,text centered]
\hspace{ 0cm}
\node[] at (1,28) {\small  \includegraphics[width=5.5cm, height=4.5cm]{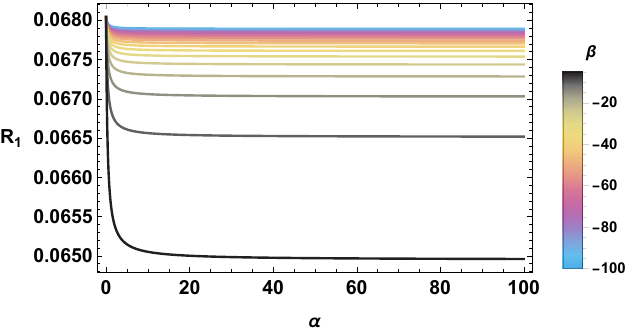}};
\node[] at (30,28) {\small  \includegraphics[width=5.5cm, height=4.5cm]{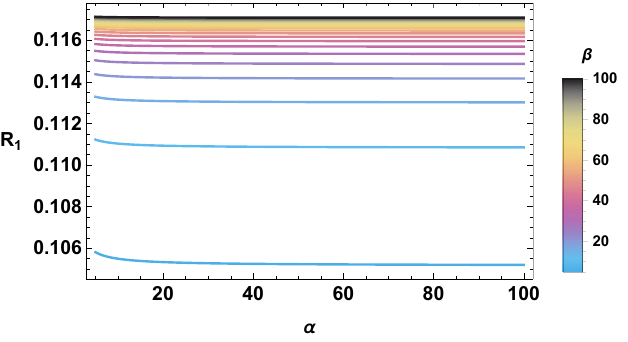}};
\node[] at (60,28) {\small  \includegraphics[width=5.5cm, height=4.5cm]{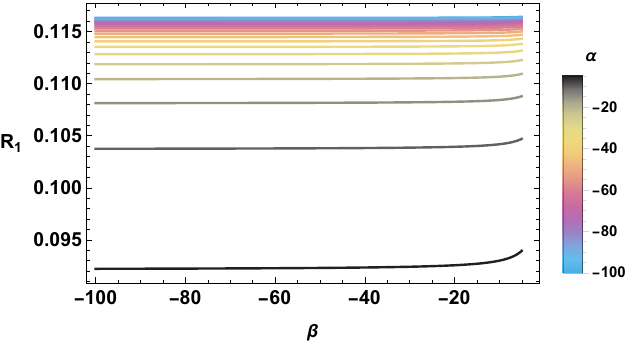}};

\node[] at (1,42) {\scriptsize    Region 1};
\node[] at (30,42) {\scriptsize    Region 2};
\node[] at (60,42) {\scriptsize    Region 3};

\end{tikzpicture}
\caption{{{\it \footnotesize Recombination factor  of the first solution  in the three allowed  magnetic charge regions. }}}
\label{bsr1}
\end{center}
\end{figure}

In Fig.(\ref{bsr1}), we depict  the recombination factor variation  in the three possible regions associated with the first  branch of solutions for  different values of the   $\alpha$ and $\beta$ parameters in  the allowed  magnetic charge regions. 

It follows from these graphical representations  that  the recombination factor    remains  less than one  in the all three  regions. In such regions,  the black strings  are stable states.   These brane objects   do enjoy the  recombination  process.

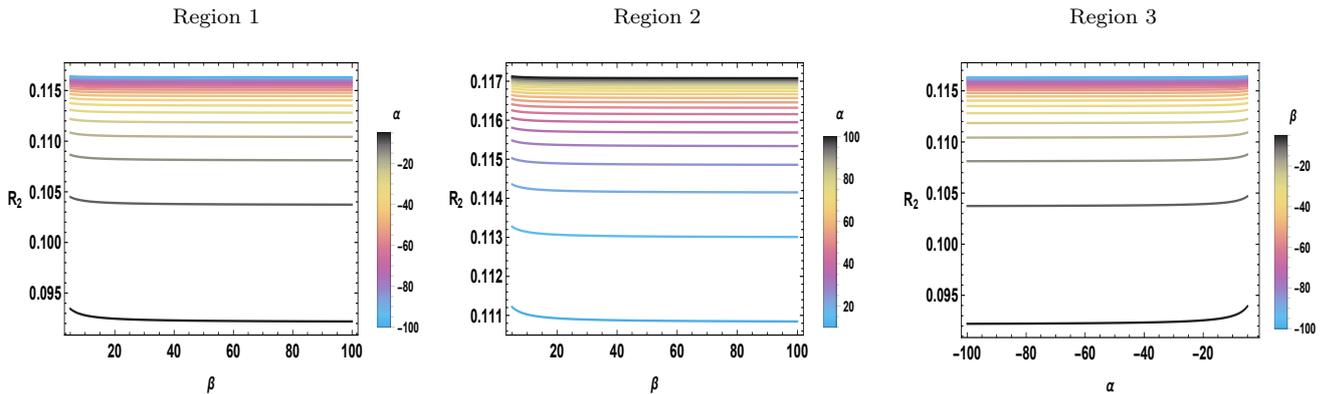
\begin{figure}[h!]
\begin{center}
\begin{tikzpicture}[scale=0.2,text centered]
\hspace{ 0cm}
\node[] at (1,28) {\small  \includegraphics[width=5.5cm, height=4.5cm]{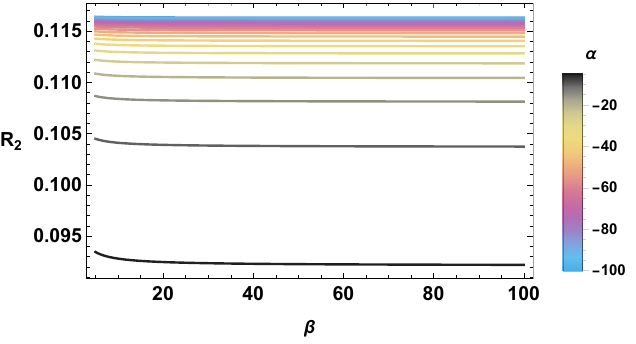}};
\node[] at (30,28) {\small  \includegraphics[width=5.5cm, height=4.5cm]{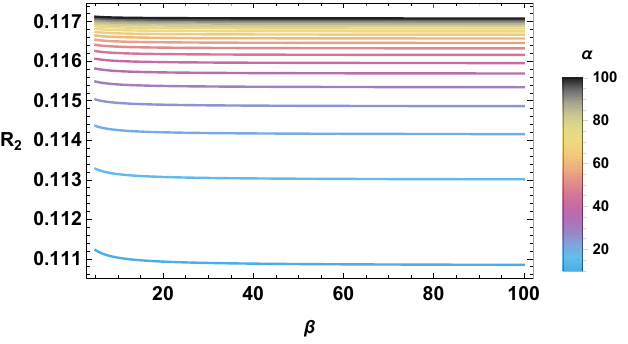}};
\node[] at (60,28) {\small  \includegraphics[width=5.5cm, height=4.5cm]{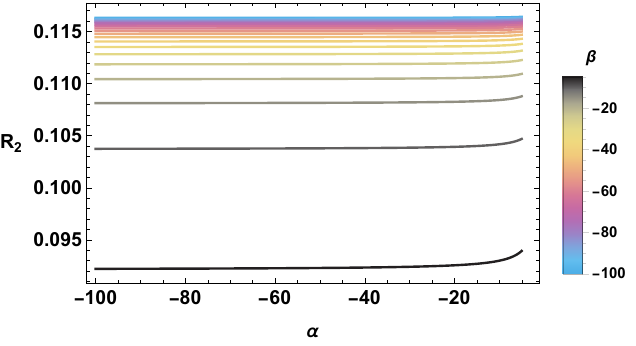}};

\node[] at (1,42) {\scriptsize    Region 1};
\node[] at (30,42) {\scriptsize    Region 2};
\node[] at (60,42) {\scriptsize    Region 3};

\end{tikzpicture}
\caption{{\footnotesize {\it Behavior of  of the second solution  in the three allowed  magnetic charge regions. }}}
\label{bsr2}
\end{center}
\end{figure}

 By  varying  $\alpha$ and $\beta$ parameters  in  the allowed  magnetic charge regions, the  values of the  recombination factor  of the  second branch of solutions are  illustrated in Fig.(\ref{bsr2}).
In  the all  regions,   we observe that the recombination factor  is always  less than one  in the allowed range of the  magnetic charge
ratios.  This shows  that the  corresponding black strings solutions  are stable. These states   enjoy  the recombination of the brane/anti-brane objects.

\section{Conclusion and discussions}
In this paper, we  have  reconsidered  the  investigation  of the   black holes and the black  strings   using the  compactification of M-theory 
on  	  a  three parameter   CY  three-fold,  constructed  as   a    complete intersection   CICY  manifold with $h^{1,1}=3$.   Exploiting   the   5D  ${ \cal N} = 2$  supergravity formalism,    we have   investigated  the 
  BPS  and non-BPS states   by wrapping M-branes  on appropriate  non-homomorphic   cycles in  the proposed   CY three-fold.    In the first part,   we have discussed  the  allowed electric charge regions of  the  BPS and the  non-BPS  black hole states from the  effective  scalar potential. Then,     we have  calculated  and  analyzed  the  entropy of such states.  Motivated by    extended black hole entropies \cite{OS1,OS2}, we  have approached the thermal behaviors by computing    the  corresponding temperature.  This  quantity together  with the entropy function   could be used  to   unveil   other  features associated with  the criticality and  the phase transition  aspects.     After that, we  have examined  the stability of  the non-BPS black holes by providing a general expression for    the recombination factor.  In particular, we have dealt with    three regions in the moduli  space   for each solution.  Examining such branch  solutions, we have discussed the stability behaviors in the allowed electric charge regions of the moduli space.   In such  allowed ranges, we have  found  stable and unstable    non-BPS states  depending on the   electric charge ratios.
   
In the second part,    we have     studied the    non-BPS  black strings  obtained from M5-branes wrapping dual non-homormphic  4-cycles in the proposed economical model.    Precisely,  we have found  multiple non-BPS solutions associated with the    three parameter  complete intersection   CICY  geometry.  Using the   extremization  mechanism with respect  to the  geometric local variables subject to the volume constraint,  we have obtained  black  string solutions from the effective scalar potential.  For such solutions, we have elaborated  a general expression  for 
   the recombination factor. After a close examination, we have shown  that the associated  non-BPS black  string states are stable in the  allowed  magnetic charge regions. In such regions, the  5D  black strings   do enjoy the  recombination process. 
  
  This work comes up with many open questions.  One  natural question concerns  extended   models associated with   higher dimensional   K\"{a}hler  moduli spaces.   We believe this will necessitate the use of high numerical computations.   We hope  to  address   such a  question in future works.

\textbf{Acknowledgement}:   The  authors   would like to thank Nour Eddine  Askour,  Saad Eddine  Baddis,  Mohamed Benali,  Chayma El Asbihani, Anas El Balali,  Wijdane El Hadri,   Hasan El Moumni,   Yassine Hassouni, Khalil  Loukhssami,  Mohamed Oualaid, Mohamed Amin Rbah, Moulay  Brahim Sedra and Yassine Sekhmani    for discussions  and collaborations  on related topics.

\end{document}